\documentclass[lettersize,journal]{IEEEtran}
\usepackage{amsmath,amsfonts}
\usepackage{algorithmic}
\usepackage{algorithm}
\usepackage{array}
\usepackage[caption=false,font=normalsize,labelfont=sf,textfont=sf]{subfig}
\usepackage{textcomp}
\usepackage{stfloats}
\usepackage{url}
\usepackage{verbatim}
\usepackage{graphicx}
\usepackage{cite}
\usepackage{hyperref}
\usepackage{booktabs}

\begin{document}

\title{The Future of Internet of Things and Multimodal Language Models in 6G Networks: Opportunities and Challenges}

\author{Abdelrahman Soliman, University of Guelph\\ asolim01@uoguelph.ca}



\maketitle

\begin{abstract}
Based on recent trends in artificial intelligence and IoT research. The cooperative potential of integrating the Internet of Things (IoT) and Multimodal Language Models (MLLMs) is presented in this survey paper for future 6G systems. It focuses on the applications of this integration in different fields, such as healthcare, agriculture, and smart cities, and investigates the four pillars of IoT integration, such as sensors, communication, processing, and security. The paper provides a comprehensive description of IoT and MLLM technologies and applications, addresses the role of multimodality in each pillar, and concludes with an overview of the most significant challenges and directions for future research. The general survey is a roadmap for researchers interested in tracing the application areas of MLLMs and IoT, highlighting the potential and challenges in this rapidly growing field. The survey recognizes the need to deal with data availability, computational expense, privacy, and real-time processing to harness the complete potential of IoT, MLLM, and 6G technology.
\end{abstract}


\section{INTRODUCTION}
\label{sec:intro}

The Internet of Things (IoT) started in 1999 when Kevin Ashton introduced the idea \cite{mouha2021internet}. Connect everyday objects to the Internet, adding smart features to homes, factories, and healthcare. IoT can greatly benefit the global economy, but it also brings risks such as security issues, privacy concerns, and moral questions about surveillance. 
A diverse array of corporations and research organizations have projected various expectations regarding the anticipated influence of the Internet of Things (IoT) on both the Internet and the global economy throughout the upcoming decade. According to \cite{manyika2015internet}, an estimated 100 billion IoT connections will be established by 2025. They also predict that the potential economic impact attributed to IoT could reach as much as 11 trillion dollars annually by 2025.

Based on the work conducted in \cite{rehman2017future}, IoT's present issues are multifaceted in nature and efficiently discourage its optimal potential. One of the main hindrances is that there are no universal standards to discourage smooth interaction and compatibility among various IoT devices and platforms. Scalability and connection stability are also serious drawbacks, as an increasing number of connected devices cause unbearable pressure on networking, addressing protocols, and routing techniques, mainly in resource-demanding devices. In addition, security and privacy are of utmost importance, with robust controls necessary to protect sensitive data generated by billions of devices from cyber threats and unauthorized access.

Following the current research trend in large language models (LLMs) with a huge adaptation from normal consumers, LLMs and IoT have been extensively discussed in the literature \cite{long20246g,lin2023pushing,ho2024remoni,sarhaddi2025llms}. Moving beyond text and reasoning, some papers tried to look at multimodality by aligning different types of data, such as images, audio, and raw bits, in the LLM framework \cite{weng2025lgvlm
,delsi2025interfacing,ho2024remoni}. In general, the main focus of the research work addressed the feasibility and application sides, while others predicted the future and suggested some future prospects and challenges.

This survey explores the potential for synergistic integration of IoT and Multimodal Large Language Models (MLLM), especially in the coming 6G networks. The survey examines how these technologies are applied in different fields, such as healthcare, agriculture, and smart cities. In addition, this work will look at the four pillars of this integration of the IoT, which are sensors, communication, processing, and security. The survey not only reveals the potential of the enabling technologies as we journey into this new era of analysis but also maps the challenges and direction for further research and development. This comprehensive overview is a handy guide for researchers wishing to expand MLLMs and IoT application domains.

\begin{figure*}[h!]
	\centering
	\includegraphics[width=1\linewidth]{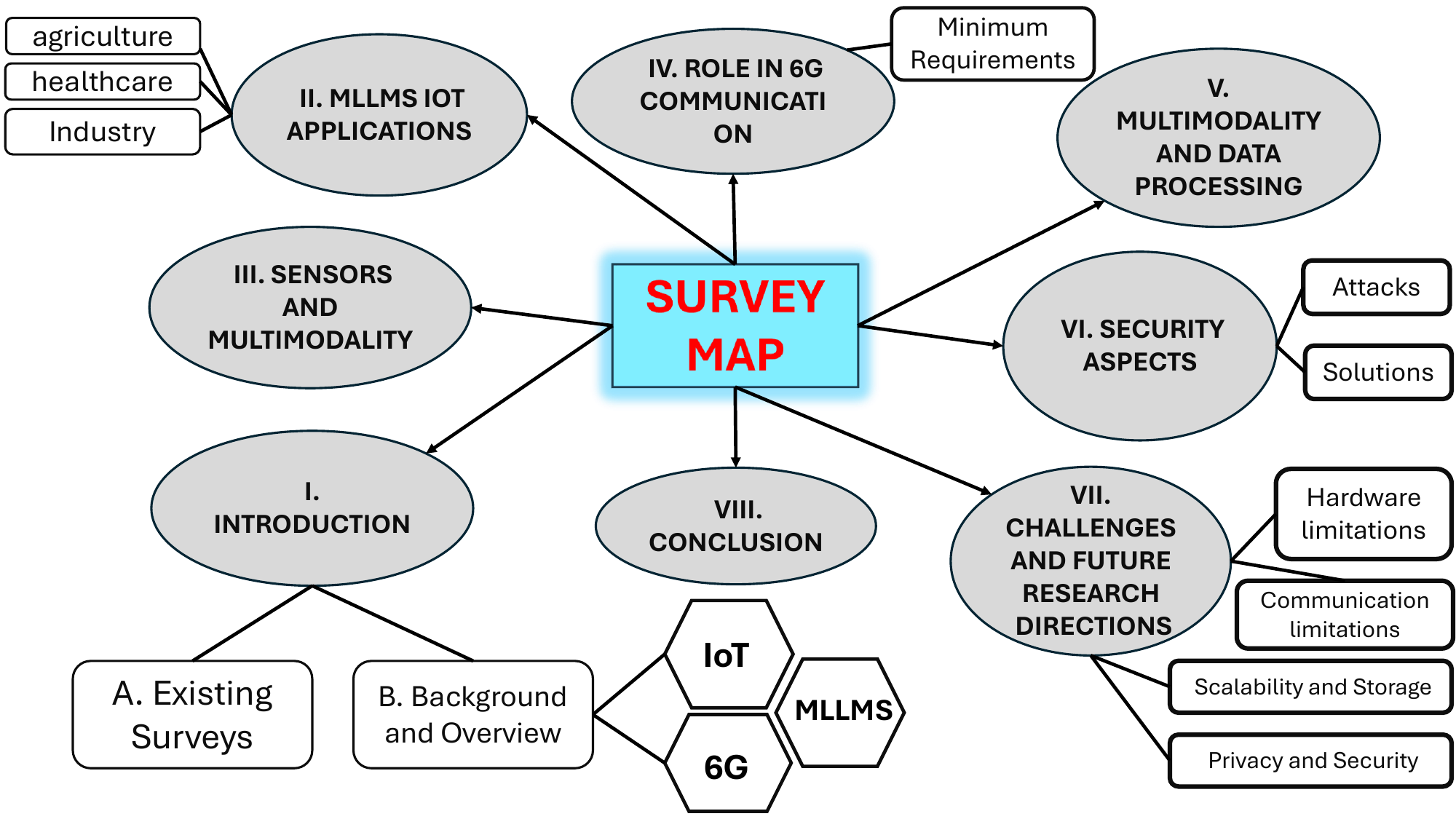}
	\caption{General Overview taxonomy of the survey}
	\label{fig:tax}
\end{figure*}

Figure \ref{fig:tax}, shows a taxonomy of this survey presenting various sections and categories.
The key contributions of this survey are listed as follows:
\begin{itemize}
  \item Detailed overview of each technology: IoT and MLLMs and their applications 
  \item Review the multimodality in four main points: sensors, communication, processing, and security.
  \item Challenges and Future Work: The open challenges and possible future directions for researchers are presented.  
\end{itemize}

\begin{table*}[h!]
\renewcommand{\arraystretch}{1.5}
\centering
\begin{tabular}{|c|c|c|c|c|c|c|c|}
\hline
Refs & IoT & LLM & 6G & Multimodality & Processing & Security & Challenges \\
\hline
\cite{boateng2024survey} & X  & \checkmark & \checkmark  & X & X         & X       & \checkmark \\
\hline
\cite{nabi2023review} & \checkmark & X  & X  & X & X    & X       & \checkmark \\
\hline
\cite{zhang2024mm} & X  & \checkmark & X  & \checkmark     & X         & X    & X         \\
\hline
\cite{qu2025mobile} & X  & \checkmark & X  & X   & \checkmark     & X       & X    \\
\hline
\cite{caffagni2024revolution} & X  & \checkmark & X  & \checkmark   & X         & X       & X         \\
\hline
\cite{sarhaddi2025llms} & \checkmark & \checkmark & X  & X      & X    & X       & X         \\
\hline
\cite{huang2024large} & X  & \checkmark & X  & \checkmark    & X    & X       & X         \\
\hline
\cite{hang2024large} & X  & \checkmark & \checkmark & X     & X         & X       & X         \\
\hline
\cite{sapkota2024multi} & X  & \checkmark & X  & \checkmark     & X      & X       & X         \\
\hline
\cite{song2023bridge} & X  & \checkmark & X  & \checkmark   & \checkmark    & X       & X         \\
\hline
\cite{kim2021multimodal} & \checkmark & X  & X  & \checkmark    & X      & X       & X         \\
\hline
\cite{zhou2025edge} & \checkmark & X  & \checkmark & X    & \checkmark     & X       & X    \\
\hline
\cite{lamaakal2025tiny} & \checkmark & \checkmark & X  & X  & \checkmark  & X       & \checkmark         \\
\hline
\cite{alsaif2024multimodal} & X  & \checkmark & X  & \checkmark      & X         & X       & X         \\
\hline
\textbf{This Work} & \checkmark & \checkmark& \checkmark & \checkmark & \checkmark    & \checkmark  & \checkmark \\
\hline
\end{tabular}
\caption{Comparison with related surveys.}
\label{tab:related}
\end{table*}

\subsection{Existing Surveys}

Researchers have published several review papers for different applications of IoT, LLMs, and 6G networks to outline how these technologies can enhance multiple services, often even providing roadmaps for forthcoming research. Although review papers across these areas mainly emphasize separate components, their in-depth interaction and convergence have yet to be touched upon. Hence, the necessary context of IoT, Multimodal Language Models, and 6G harmoniously interacting is absent. For example, the authors in \cite{sarhaddi2025llms,dou2023towards,zhou2025edge} have published comprehensive review articles on topics of LLM and IoT, AGI in IoT, and Edge Intelligence for 5G and IoT. Although the work in \cite{sarhaddi2025llms} reviews the domain of LLMs and IoT, it does not delve into specific networks or multimodal roles. Similarly, \cite{qu2025mobile} investigated the intelligence of the mobile edge in large language models, but their work does not specifically highlight the field of IoT or the general synergy with the progress of 6G communication and multimodal data. However, work \cite{ghimire2025enhancing} concentrated on the improvement of security in IoT systems employing LLMs while offering useful points regarding cybersecurity without extensively investigating the possibilities facilitated through 6G and multimodality.

The authors in \cite{hang2024large} reviewed large language models in next-generation networking, such as 6G, but not exactly on the revolutionary impact in the context of IoT and the utilization of multimodal data. Furthermore, we did not find any survey articles that addressed the collective potential of multimodal language models in the 6G-enabled IoT paradigm. \cite{caffagni2024revolution} reviewed multimodal large language models and, while comprehensive in MLLM architectures, does not specifically target the IoT and 6G application space. 

However, in contrast to available surveys, our survey is unique because it provides an integrated and overall perspective of the converging paradigm of the Internet of Things, Multimodal Language Models, and 6G networks. We specifically discuss the potential and challenges through such convergence in a multi-odality with sensors, communication, processing, and security. We offer a comprehensive review of ongoing research activities that aim to leverage MLLMs in 6G-IoT, describe the primary challenges and probable future research directions, and present these based on our comprehensive review. Table~\ref{tab:related} briefly compares our work with the most relevant surveys in the literature, highlighting our unique focus on the synergy of IoT, MLLMs, and 6G.

\subsection{Background and Overview of technologies}

\begin{figure}[h]
	\centering
	\includegraphics[width=0.97\linewidth]{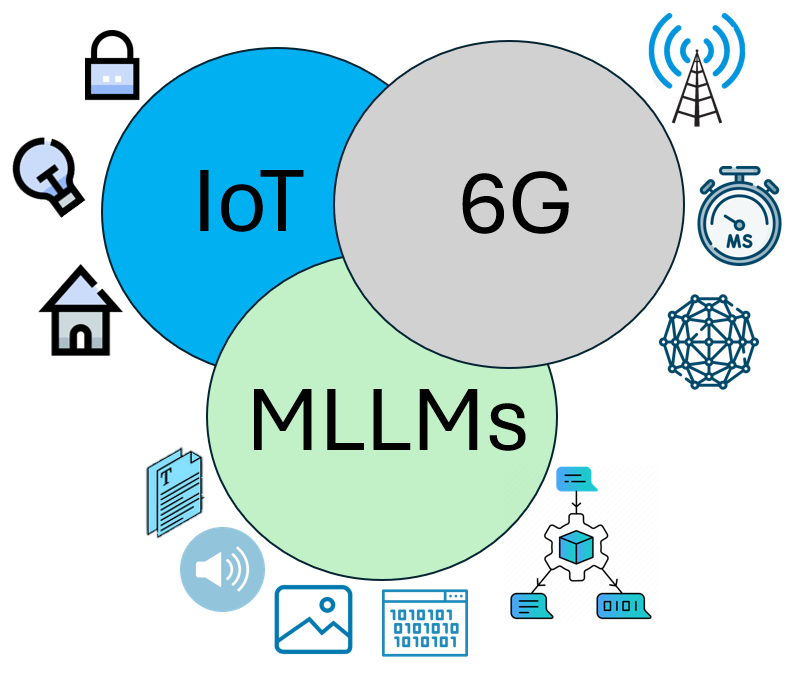}
	\caption{Overview of main topics}
	\label{fig:Background}
\end{figure}

As shown in Figure \ref{fig:Background}, this section provides an overview and background on the three main topics that will be discussed in this survey, including IoT, 6G, and MLLM.

\subsubsection{IoT}
The Internet of Things (IoT) is defined as a network of physical objects, "things," embedded with sensors, software, and other technologies to connect and exchange data with other devices and systems over the Internet. There are various complementary components in an IoT system. Sensors act as an interface to the world for the system, sense and measure world states, and interpret them as digital data. Actuators respond to data processing by making physical impacts on the environment, such as interaction and control. Communication networks on top of divergent range technology from low to wide area make transfer of information across devices, gateway, and the cloud. Edge and cloud-provided distributed processing takes the unprocessed sensory inputs and maps them into insight with different algorithms applied. Finally, robust security measures are built at all points, from the device to the app layer, to provide robust protection from hazards and pitfalls throughout the overall system and for information.

IoT faces many challenges and requires new technology or frameworks to manage the expansion of the number of devices and demands. Limited computing resources and the requirement of real-time reaction in most IoT devices, often sized and power-constrained, restrict the level of on-device processing complexity and require data efficiency handling for effective action \cite{dou2023towards}. High-density IoT communication networks with massive device density and heterogeneity of communication protocols present problems such as network administration, data congestion, and scalability \cite{boateng2024survey}. Furthermore, security and privacy are the matters of most concern given the high degree of sensitive information harvested by IoT devices and the interconnectedness of devices that create extensive exposure to cyberattacks and invasion of privacy and require robust and participatory security measures \cite{ghimire2025enhancing}.

\subsubsection{6G}
The research community has now shifted its focus toward developing 6G wireless systems to overcome the limitations of current 5G technology, such as bandwidth and latency constraints, particularly for new applications such as the metaverse and dense IoT systems. Various organizations around the world, from universities to industries, have defined their vision for 6G wireless and have initiated related research activities accordingly \cite{wu20216g}. 

Until 2030, the hope is that 6G will fully exploit its potential in the areas of energy and environmental sustainability, digital accessibility, and industry flexibility, such as in health and safety. As presented in Table \ref{tab:5G_6G_comparison}, the main goals of the 6G network are to provide high data rates, energy efficiency, global coverage, and intelligent and secure connections \cite{quy2023innovative}.

To start with, 6G is poised to offer high data rates, where it will be possible to deliver up to 1 terabit per second (Tbps). This is much larger than the current 5G systems, with faster data transfer and support for new use cases requiring high bandwidth. Secondly, it provides high energy efficiency for limited-resource personal mobile devices. As more mobile devices are added and the demand for wireless connectivity increases, energy efficiency has emerged as a key issue. The purpose of 6G systems is to address this problem by providing energy-efficient solutions to facilitate the long-term sustainability of wireless communications \cite{jiang2021road}.

In addition, the next generation will construct a global coverage network consisting of space, air, ground, and underwater parts. The end-to-end network infrastructure will enable seamless connection and end-to-end services regardless of the user's location and environment. Finally, 6G focuses on providing intelligent and reliable connections throughout the network. With the use of emerging technologies such as artificial intelligence (AI) and machine learning (ML), 6G networks will enhance the performance of the network, as well as enhance the user experience and ensure reliable connectivity in various contexts.
From a service class perspective, the transition is expected to evolve from ultra-reliable low-latency communications (URLLC) and enhanced mobile broadband (eMBB) to novel enhanced services,
such as massive URLLC (mURLLC) and reliable low-latency mobile broadband communication (mBRLLC)  \cite{zhang2021age}. 

\begin{table}[h]
\renewcommand{\arraystretch}{1.9}
\centering
\begin{tabular}{|c|c|c|}
\hline
\textbf{Technology} & \textbf{5G} & \textbf{6G} \\
\hline
\textbf{Applications} & eMBB, URLLC, mMTC & AI-IoT. AR, XR, VR \\
\hline
\textbf{Data rate} & 10 Gbps & 1 Tbps \\
\hline
\textbf{Frequency} & 3-300 GHz & 1000 GHz \\
\hline
\textbf{Latency} & 10 ms & $<$1 ms \\
\hline
\end{tabular}
\caption{Comparison of 5G and 6G Technologies.}
\label{tab:5G_6G_comparison}
\end{table}

\subsubsection{MLLMs}
The appearance of the transformer model in 2017 by \cite{vaswani2017attention}, "Attention is All You Need." It was a breakthrough after years of sequence-to-sequence research that was largely dominated by recurrent neural networks (RNNs) and long short-term memory networks. The transformers address some shortcomings of the earlier models in dealing with sequences. For one, parallelization is possible with them as full sequences are processed in parallel, which vastly improves training and inference speed over RNNs. The self-attention mechanism facilitates the establishment of long-range dependencies within sequences. Secondly, it allows better scalability and, therefore, can handle longer sequences and larger datasets more gracefully than RNNs. Thirdly, their architecture is general, and therefore, it can facilitate numerous tasks beyond Natural Language Processing.

\begin{figure}[h]
	\centering
	\includegraphics[width=0.97\linewidth]{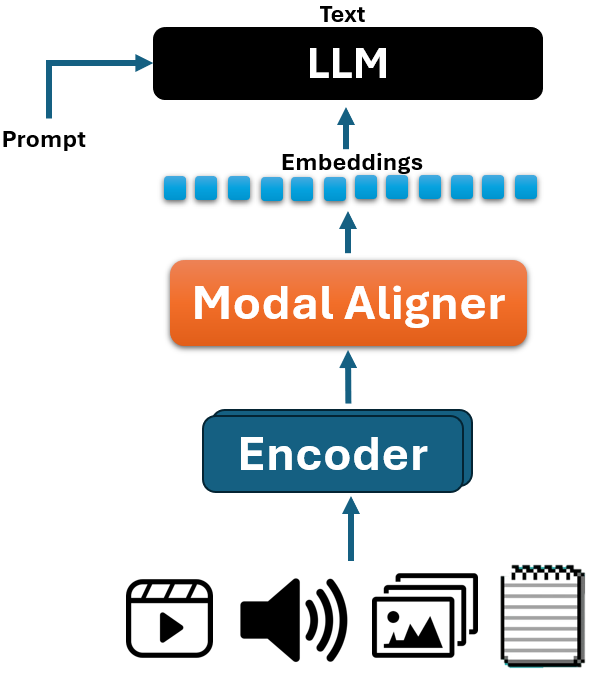}
	\caption{Multimodal LLMs architecture}
	\label{fig:mllms}
\end{figure}

Large language models (LLMs) are based on transformer architecture. Specializing in text and summarization.
The evolution of LLMs has opened the door to the potentially revolutionary innovation of Large Multimodal Models (LMMs) or Multimodal Large Language Models (MLLMs), which extend the capabilities of language models to understand and generate content across various modalities, such as images, speech, raw bits, and video \cite{huang2024large}. As shown in Figure \ref{fig:mllms}, LMMs do this by implementing methods such as multimodal encoders to handle multiple inputs, modal aligners to align the various feature spaces, and multi-modal decoders to produce output outside text. Despite huge advances, LMMs still have limitations in bringing multimodal comprehension together, sustaining coherence between modalities, and keeping computational costs in check with increasing model scales. The direction of future LMM research is to develop more effective architectures, improve prompt engineering for multimodal input, and explore new training methods to produce truly adaptive and robust multimodal AI systems that can handle subtle real-world scenarios \cite{cheng2025large}.

\begin{figure*}[h!]
	\centering
	\includegraphics[width=1\linewidth]{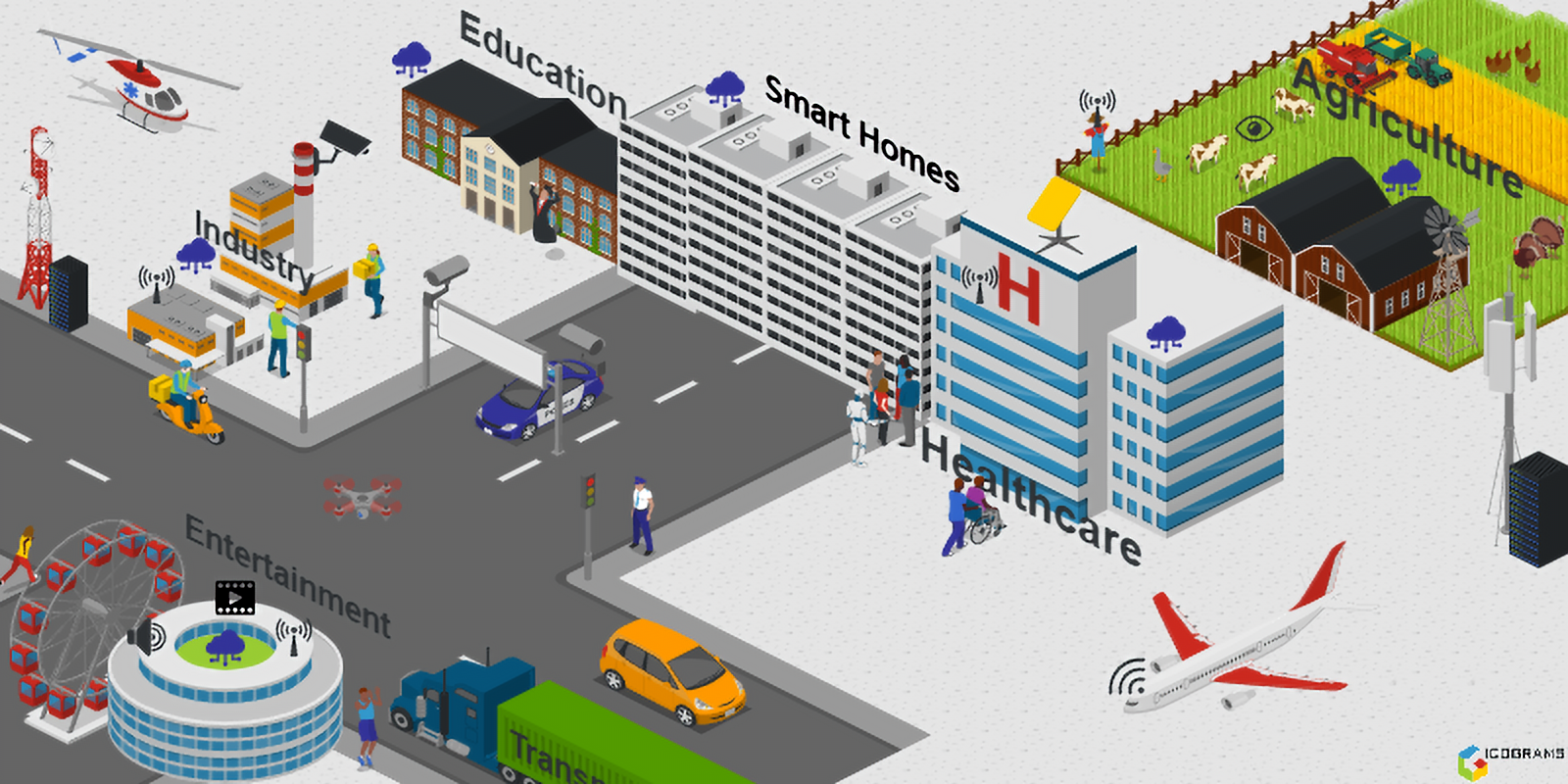}
	\caption{MLLMs Applications}
	\label{fig:mllms_application}
\end{figure*}

The embedding is crucial for projecting heterogeneous forms of data into a shared representation upon which the model acts. In visual data like images and videos, Convolutional Neural Networks (CNNs) and Transformers extract the visual features from techniques like convolutions and attention mechanisms. Alternatively, in the audio-text case, recurring neural networks (RNNs) and transformers extract sequential information and interrelations. Modality-specific depth, thermal, and IMU information have gained domain-specific neural networks for temporal and geometric dynamics. Finally, the result of the above embedding processes is a shared embedding space where the characteristics of various modalities can be compared and fused relatively in a balanced manner \cite{tharayil2025multimodal}.

\section{MLLMS IoT APPLICATIONS}
\label{sec:applications}

The integration of IoT, MLLMs, and 6G offers a variety of services and applications in different fields. As shown in Figure \ref{fig:mllms_application}, whether it is health, smart cities, or education, it is conjectured that these applications and devices will be everywhere in our lives.

For healthcare applications, the authors of \cite{ho2024remoni} propose advanced remote health monitoring through a seamless pairing of multimodal large language models (MLLM) with heterogeneous data streams. The breakthrough with the system lies in the integration and processing of information from varying modalities: accelerometer and vital signs data obtained from wearable Internet of Things sensors and visual information obtained from patient video samples. This multimodal design enables the MLLM-based natural language processing module to extend beyond common vital sign monitoring and recognize and understand fine-grained patient activity and emotional states from video in a direct manner. In addition, prompt engineering methods play a substantial role in keeping these heterogeneous data modalities in balance so that there is still a coherent and integrated representation of the patient's state.

Driven by their superior generalization and reasoning abilities compared to traditional AI models. LLMs and Vision Language Models (VLMs) such as Med-PaLM demonstrate the potential for LLMs to function as AI medical generalists, capable of high-quality medical inquiry responses and even exceeding professional benchmarks such as X-ray diagnostics. The concept of personalized health AI experts is becoming more realistic, offering diverse services from chatbots to early warnings through multimodal inputs and outputs. However, cloud-based deployment of such healthcare LLMs faces significant hurdles, including massive multimodal data transmission demands and, critically, patient data privacy concerns and stringent regulations that impede centralized training \cite{lin2023pushing}. Similarly, in robotics, MLLMs like PALM-E \cite{driess2023palm} empower embodied reasoning and complex task planning directly from raw sensor data, yet centralized training models for robotics face challenges of overwhelming video data uploads, sensitive interactive data privacy, and stringent low-latency requirements for real-time control, especially in human- machine interaction scenarios. 

In agriculture, LLMs are aggressively sought after in agriculture for an array of uses, specifically multi-modal LLMs that have much promise \cite{li2023large}. Models are very capable of making complex agricultural issues easy to tackle and improving decision-making by analyzing heterogeneous agricultural data such as images, text, and sensor data. Applications span crop monitoring, disease diagnosis, precision irrigation, and image processing in agriculture. Integration with LLMs comes with challenges. The significant drawbacks are data availability and quality, in particular, the need for high-quality domain-specific training data. In addition, the high computational costs and complexity involved in integrating LLMs in existing agricultural systems are still major challenges to be addressed for their effective and large-scale implementation in future agricultural systems.

With applications spanning from data integration to decision support. LLMs and MLLMs can be used to power IoT and Sensor Data Integration, which would enable more precise data collection and analysis to facilitate real-time farm condition monitoring \cite{sapkota2024multi}. Such data powers advanced predictive crop analytics to enable precise yield predictions and optimize resource utilization. In addition, these models should improve automated crop management systems capable of maximizing irrigation and fertilization policies. Moreover, AI-based pest and Disease Detection is another potential area where VLMs maximize image processing to detect early and accurate diagnoses \cite{wu2023extended}. One step beyond the farm gate, LLMs can assist in agricultural supply chain optimization, streamlining production-to-market processes \cite{darapaneni2022lstm}.

This breakthrough in MLLMs and IoT devices with fast and reliable 6G communication will enable fully connected smart cities, aiming to simplify the lives of residents. According to \cite{tami2024usingmultimodallargelanguage}, MLLMs could be leveraged to control critical traffic using text and image modalities. In the same direction, work in \cite{cui2024drive} proposed a framework for the control of autonomous vehicles using the multiple modalities and sensing capabilities of MLLM along with different IoT sensor inputs.

As an insight obtained through previous uses, the merging of IoT, MLLMs, and 6G presents a good framework for general use cases, calling on multimodal data for better intelligence and automata in other scenarios such as industry, entertainment, and education. However, realizing this potential depends on adequate care for critical issues such as data availability and quality, computational cost, rigorous data privacy policies, and low-latency processing requirements, especially in real-time systems. Conquering these issues is the main factor in the successful and widespread use of these combined technologies.

\section{Sensors and Multimodality}
\label{sec:sensors}

Usually, IoT devices have multiple passive and active sensors, and there are design studies on planning, control of the sensors, and their location. The multimodality of these sensors means that we need to get the most information using the fusion of the data coming from each sensor. The advantages of MLLMs reside in their structure and training. As a countermeasure to standard practice, MLLMs are trained in joint training frameworks, allowing multimodal input processing in a unified framework. Joint training would presumably allow for multimodal correlation and complementarity to be learned through the model while jointly optimizing the network parameters in a manner suited to learning for global representation and understanding of data \cite{cheng2025large}.

MLLMs, as an advancement to LLMs, possess good generalizability. This is vital for integrated sensing and communication (ISAC) systems that need to operate within dynamic and covert environments, challenging traditional models tailored to specific types of data or tasks. The work in \cite{cheng2025large} presented a study of beam prediction that demonstrates this capability, as it is reported that MLLMs, under the command of the algorithm and using GPS and RGB images, perform better than traditional algorithms in the prediction of the best beam indices. Despite these advantages, there are also some serious challenges to overcome, including high computational demands on MLLMs, the need for large quantities of multimodal training data, and the complication of ensuring data privacy and security in ISAC systems.

\begin{figure}[h]
	\centering
	\includegraphics[width=1\linewidth]{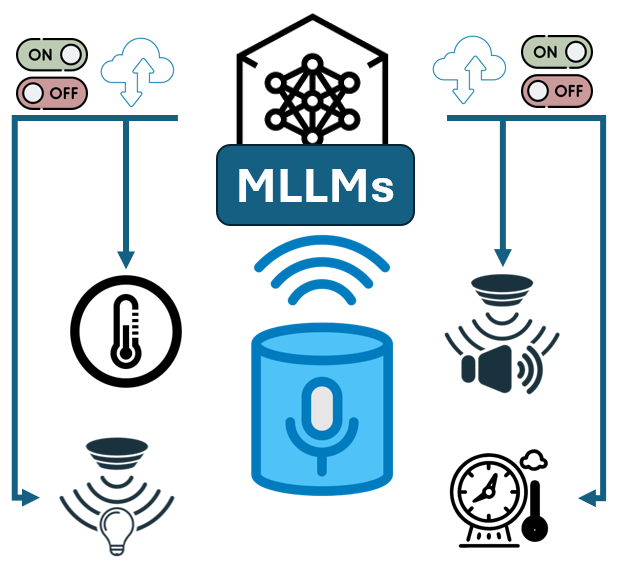}
	\caption{MLLMs and Sensors}
	\label{fig:mllms_sensors}
\end{figure}

As shown in Figure \ref{fig:mllms_sensors}, MLLMs, with their ability to understand complex multimodal contexts, can be used to offer dynamic sensor planning and control. For instance, when RGB-D cameras provide sufficient visual information under ideal lighting conditions, MLLMs would be capable of smartly downgrading the dependence or even disabling power-consuming sensors like LiDAR or radar. Alternatively, during nighttime environments, the system could prioritize radar or infrared signals, dynamically adjusting the use of the sensor according to the environment context and task requirement \cite{baris2025foundation}. Moreover, MLLMs, through multimodal instruction calibration, can be instructed to read and execute power management instructions. This could involve methods such as adaptive sensor sampling rates, which are higher when more accurate information is needed. By intelligent control of sensor power-up and runtime parameters based on real-time multimodal understanding, MLLMs offer a key to significantly enhanced power efficiency across IoT systems, supporting increased sustainable and more resource-efficient deployment \cite{tharayil2025multimodal}.

\section{Role in 6G Communication}
\label{sec:comm}
This section discusses the communication aspect of IoT, starting with the basic communication requirements for IoT systems and the role of MLLMs. 

\subsection{Minimum Requirements}
To investigate the potential of 6G in MLLM-IoT systems, the essential communication needs must be examined. These depend on three factors: architecture and processing location, data modality, and application requirements.

The location of MLLM processing plays an important role in communication. Cloud-based solutions require high bandwidth to transport raw IoT data to far-end servers. Edge computing reduces this by working in proximity to devices, transmitting processed intelligence alone, which reduces bandwidth consumption to a large extent \cite{yu2024model}. On-device processing, where feasible, would limit communication to simple commands, the absolute minimum. Positioning processing in proximity to the IoT device naturally reduces communications requirements.

The volume and type of data are also important. Text-based IoT has low communication requirements. Image and audio data require higher bandwidth requirements. Video data require the highest bandwidth requirements. Multimodal data combines them, making even higher requirements. The larger volume of data and the more sophisticated modalities translate directly into a higher minimum communication bandwidth. Compression efficiency and edge preprocessing become necessary with high volumes of data. Communication requirements are determined by application requirements. Latency-sensitive applications require low-latency communications and, more broadly, edge processing and potentially higher bandwidth for real-time data communication \cite{zhou2025edge}.

\begin{figure}[h]
	\centering
	\includegraphics[width=1\linewidth]{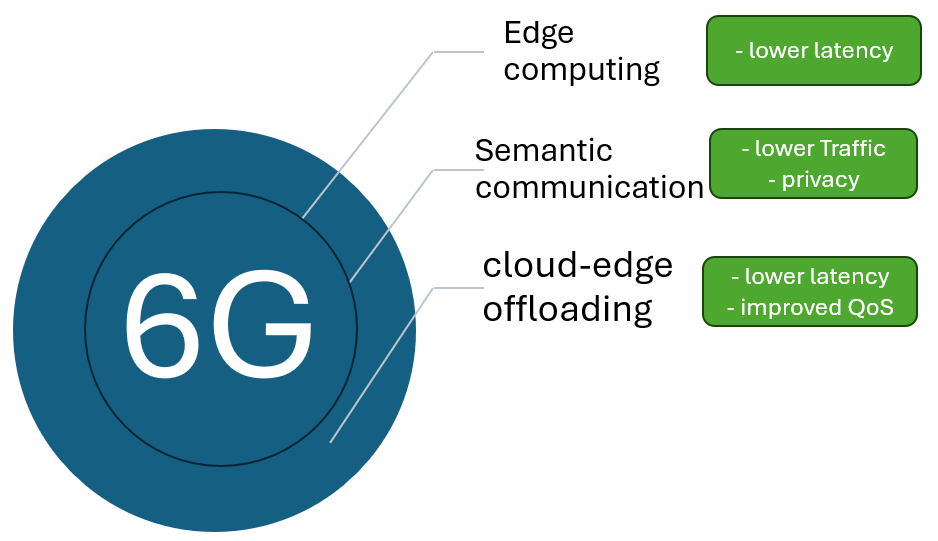}
	\caption{Role of 6G techniques using MLLMs for IoT}
	\label{fig:6g_mllms_}
\end{figure}

6G mobile edge computing and LLMs are things that can greatly contribute to the IoT environment. The localization of LLMs to the edge of 6G, as advocates \cite{lin2023pushing }, clearly resolves the main disadvantages of the current cloud-based IoT setup. For the first time, the lower latency introduced by 6G MEC is crucial in real-time IoT applications. Take the case of IoT-powered smart factories, where LLM-driven robotic arms require immediate responses to sensor inputs to move with accuracy and safety. LLM inference via the cloud adds unacceptable latency to degrade real-time control. Second, the bandwidth efficiency of the 6G along with edge computing becomes significant with IoT devices producing increasingly richer multimodal information, like images and video from intelligent sensors or robots. Centralizing such information for cloud-based LLM processing would overwhelm backhaul networks. Edge-deployed LLMs, inferred and trained closer to where data originate, limit bandwidth requirements. Finally, privacy concerns built into IoT data collection, particularly in sensitive domains such as smart homes or healthcare IoT, are handled by edge LLMs.  

Based on work \cite{boateng2024survey}, in the IoT setting, LLMs and MLLMs have the role of handling the massive amount of data generated by billions of connected devices. Through data analysis, LLMs are capable of facilitating intelligent decision making for 6G network deployment and resource allocation. For example, LLMs can optimize the use of IoT device resources, ensuring seamless resource requests and responses. The authors in \cite{cao2024multimodal} propose that the IoT will directly benefit from 6G, semantic communication, and multimodal large language models (MLLMs), as discussed in their paper. The vast volume and variety of data from IoT sensors are an issue directly addressed by 6G and semantic communication as the more efficient transmission of data and less data traffic. Semantic communication, particularly when MLLMs empower it, offers a means of transcending the limitations of traditional syntactic methods to be geared toward conveying only the essential semantic information of IoT data and not the entire raw data stream. MLLMs, being able to comprehend and process multimodal data in a homogeneous and homologous manner, become super semantic encoders and decoders for any kind of IoT sensor data (images, audio, sensor values, etc.).

This research in \cite{hu2024cloud} designs a cloud-edge collaborative system approach to Advanced Driver Assistance Systems (ADAS) in IoT environments with MLLM. The assumption is to have a domain-specific and cost-reduced MLLM maintained at the edge for rapid action and a larger and more capable MLLM (ChatGPT-40) running on the cloud to perform higher-end reasoning and context perception. The use of MLLM improves the robustness and accuracy of ADAS by allowing a more effective fusion of multisensor data and situational awareness beyond the limitations of traditional rule-based or learning-based approaches. In addition, the optimized offloading strategy improves the offloading of tasks, which results in less latency, energy conservation, and improved QoS, all of the most important factors in enabling the deployment of advanced AI-based services like ADAS over resource-constrained IoT scenarios, indirectly complementing the goals of 6G networks to offer support for more complex and demanding applications with higher efficiency and reliability.

\section{Multimodality and Data processing}
\label{sec:processing}

\begin{figure}[h]
	\centering
	\includegraphics[width=1\linewidth]{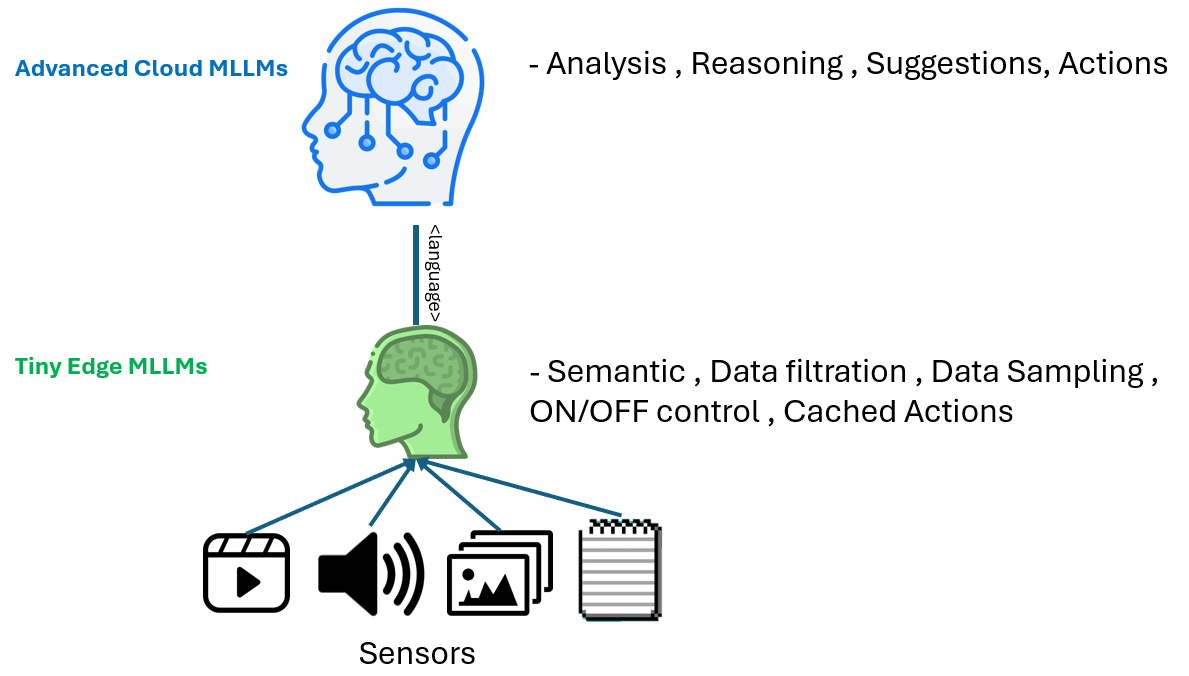}
	\caption{Multimodality and Data processing Architecture}
	\label{fig:processing_mllms}
\end{figure}

Having raw data from sensors is useless until we filter it and process it into useful knowledge and actions. Current data processing ranges from classic mapping functions to machine learning algorithms. In the realm of MLLMs-IoT, MLLMs will work as orchestrators and core analyzers of the data coming from the sensors. The work in \cite{kok2024iot} stated that MLLMs and LLMs are going to revolutionize data processing in IoT with advanced reasoning and decision-making capabilities. These are intelligent minds for the IoT, which enable them to process massive streams of sensor data, understand complex contexts, and make smarter and more sophisticated decisions. Such a two-way mutualism is similarly two-way in a concern with direction because the IoT provides LLMs with domain-based real-world facts, which counteract hallucination issues and ground their learning on real-world sensory stimuli. Conversely, LLMs, on the other hand, tackle issues of human-device interaction inferior to the best by providing natural language interfaces, processing heterogeneous data streams efficiently, and dynamically responding to ever-changing conditions. Although these are huge benefits, there are interoperability issues with IoT and LLMs. The scale and computationally intensive nature of LLM models incur huge hardware and computation requirements with immense bandwidth constraints \cite{lin2023pushing}.

As shown in Figure \ref{fig:processing_mllms}, the hierarchical structure in which the sensors at the lower level pick up heterogeneous data types. The raw sensor data are then fed through Tiny Edge MLLMs, which represent edge-level processing. Operations at this level include semantic analysis to determine the meaning of sensor data, data filtration to eliminate unwanted noise or information, and caching action to obtain a quick response \cite{qu2025mobile}. Subsequently, processed information is communicated in a language to Advanced Cloud MLLMs. In the cloud layer, higher-level processes are executed, including in-depth analysis and reasoning on the data gathered, based on the high-level realization of the situation. This multitiered approach optimally assigns processing tasks, using edge computing for local and real-time response and cloud capacity for advanced analytics and decision making.

The work in \cite{mo2024iot} introduced IOT-LM, a new large multisensory language model for the IoT setting. To facilitate this, the authors created a large-scale multilingual data set containing more than 1.15 million samples in 12 different IoT modalities and 8 tasks. IOT-LM employs a new multisensory multitask adapter layer to condition pre-trained large language models on diverse IoT sensor data, enabling the sharing of information between tasks and modalities. Experimental results demonstrate that IOT-LM significantly surpasses existing work on 8 supervised IoT classification tasks and has new strengths in interactive question-answering, reasoning, and dialog based on sensor data. 

\section{Security aspects}
\label{sec:security}
As shown in Table \ref{tab:related}, many surveys do not cover security studies of LLMs and MLLMs in IoT.
Having a robust and secure system is crucial, as many IoT applications play a pivotal role in imperative areas such as healthcare, industry, and agriculture.  LLMs and MLLMs present serious security threats due to their complexity and the finesses involved in training and use. These attacks exploit the nature of LLM and can be used in several training paradigms, including supervised learning, adversarial learning, and reinforcement learning. The multitype nature of attacks against LLMs deployed in networks necessitates a thorough understanding of such attack types to implement productive countermeasures.

\begin{figure}[h]
	\centering
	\includegraphics[width=1\linewidth]{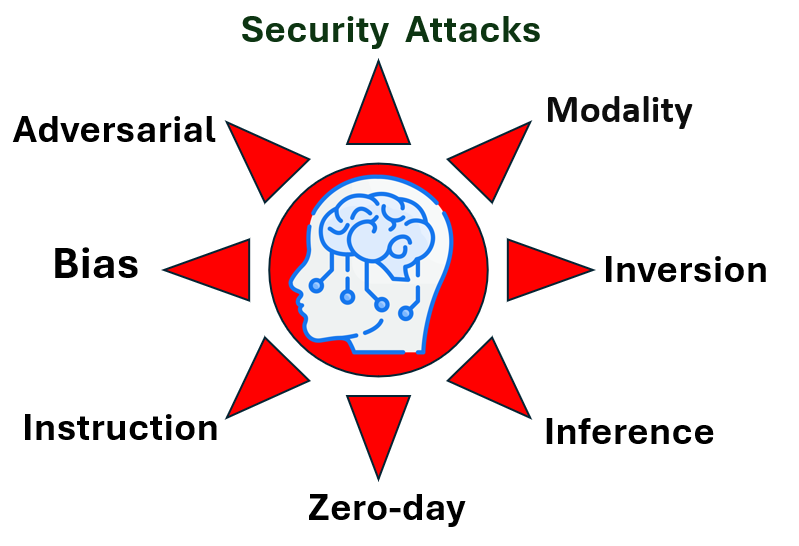}
	\caption{MLLMs Attacks}
	\label{fig:attacks_mllms}
\end{figure}

LLM attack space includes several distinguishing types, as presented in Figure \ref{fig:attacks_mllms}. Adversarial attacks mislead input information in a manner that model performance declines, as well as backdoor attacks with trigger embedding and poisoning attacks that are used for corrupting training information. Inversion attacks attempt to retrieve training data or information stealing from model gradients, leading to model copying or data theft. Unfair exploitation and bias attacks employ biased training data to spread disinformation and negative biases. Instruction-tuning attacks attempt to drain system resources, including Denial of Service, prompt injection, and jailbreak attacks. A modality attack means that attackers might craft inputs where different modalities present conflicting or ambiguous information. This could cause the MLLM to become confused, hallucinate information, and produce incorrect output as it struggles to understand the modalities. Zero-day attacks or sleeper agents are dormant in model weights, waiting for specific events to carry out activities like data exfiltration. Finally, inference attacks, namely membership inference, try to obtain sensitive data about training data, potentially breaching confidentiality and user privacy \cite{khowaja2024pathway}.

The generative and analytical power of LLM enables it to be applied to solving a broad variety of network security scenarios. On average, LLMs can be trained in normal network traffic, user activity, and system log patterns. LLMs can be detected for anomalies by comparing new data with those already known \cite{guo2024large}. The authors in \cite{ferrag2024revolutionizingcyberthreatdetection} presented SecurityBERT based on the bidirectional encoder representations of the Transformers (BERT) model to detect network threats in IoT networks. During training, it employs a novel privacy-preserving encoding technique to encode network traffic data succinctly in a structured manner. In addition, LLMs can analyze software code, system configurations, and network structures to identify potential security vulnerabilities. Using various prompt technologies, LLMs would achieve software vulnerability detection.

The work in \cite{ghimire2025enhancing} proposes a hybrid cybersecurity solution for critical infrastructure that combines auto-encoders for numeric anomaly detection LLMs for enhanced security and interpretability. The results demonstrated that pre-processing with LLM significantly improved anomaly detection performance, from a macro-average F1 score of 0.49 to 0.98. Furthermore, the system uses GPT-4 to generate natural language explanations of detected anomalies, providing context and enhancing the interpretability of the system for cyber analysts.

Another work focused on a proactive intrusion prediction model for IoT systems powered by
pre-trained LLM. The model employs two LLMs: a fine-tuned Bidirectional and AutoRegressive
Transformers (BART) model for network traffic prediction and a fine-tuned Bidirectional Encoder Representations from the Transformers (BERT) model for testing the predicted traffic. Using the CICIoT2023 IoT attack dataset, the model shows a remarkable improvement in predictive performance, with an accuracy of 98\%.

Regarding multimodality and security, the paper \cite{jiao2025enhancing} focused on the promising application of LMMs to improve the security of IoMT networks. Knowing that IoMTs are growing in complexity and have a high data volume, the authors propose that a multimodal model improves anomaly detection through more accurate traffic prediction, enabling adaptive response to changing cyber threats and improving physical layer security (PLS). Solutions for PLS improvement are aimed at better fusion techniques like attention mechanisms and Graph Neural Networks (GNNs) to support heterogeneous signals, as well as reinforcement learning for optimization and edge-based inference for real-time inference under dynamic IoMT scenarios.

\section{CHALLENGES AND FUTURE RESEARCH DIRECTIONS}
\label{sec:challenges_future}
This section will dive into the current challenges that limit the applications of MLLM and IoT on 6G networks. Finally, future directions will be presented for researchers to carry on with the next generation of IoT. 

\subsection{Technological challenges}
\begin{enumerate}
\item \textbf{Hardware limitations:}
One major challenge to LLM integration is the hardware constraints of IoT devices, which cannot cope with their enormous model sizes, and many IoT devices have limited power and storage volumes. Running complex computations locally while ensuring efficient resource utilization can be a significant challenge. Strategies such as model compression, optimization, partitioning, and resource management provide some respite by minimizing computational requirements. However, there is still much work to do, especially for low-resource devices \cite{kok2024iot}. Existing solutions, such as edge- cloud collaborative systems and hierarchical LLM chains, often bring in latency problems that detract from real-time performance. This hardware limitation also probably explains the limited application of advanced reasoning algorithms on edge devices.

\item \textbf{Scalability and Storage:}
The storage of gigantic real-time streams of data from networked IoT devices and sensors is a serious concern. Data that are huge, varied, fast-moving, and often have to be processed in real time require high-availability storage facilities. Data that grows exponentially with increasing IoT ecosystems can easily flood traditional storage systems \cite{dou2023towards}. In addition, the nature of IoT data, being structured, semi-structured, or unstructured, demands elastic storage facilities. The inability of low-cost sensors to manage data variety and scalability also constrains the possible scale of IoT deployments.

\item \textbf{Communication limitations:}
As more and more IoT devices become part of our daily lives, there is an enormous communication challenge. With an increasing number of devices comes massive interference in how they can communicate with each other \cite{dou2023towards}. This is particularly clear with wireless communication, where the resources to transmit signals are limited, and increasing the resources or spectrum is not a viable solution. Consequently, much effort is being directed toward creating better ways of communicating using the reasoning capabilities of LLMs and semantic communication \cite{wang2022machine}.

\end{enumerate}

\subsection{Other challenges}
\begin{enumerate}
    \item \textbf{Privacy and Security:} Although Federated Learning (FL) offers privacy gains for Large Language Models (LLM), the current literature demonstrates that their privacy can be violated by model weights \cite{lin2023pushing}. Differential privacy (DP) is a stronger approach to protecting privacy with formal privacy assurances. Existing IoT-LLM privacy techniques, like adaptive differential privacy and federated learning, are cloud-dependent and thus come with latency and expose subject data to transmission vulnerabilities. In order to address this, future research must develop fully embedded lightweight LLMs for edge devices without dependence on the cloud and loss of performance \cite{kok2024iot}. 

    \item \textbf{Legal Considerations:} The government and the organization needed to establish regulations and user data access to prevent misuse and side effects on the output of these multimodal models in critical applications such as healthcare \cite{nabi2023review}. 
\end{enumerate}
\subsection{Future research directions}
\begin{enumerate}
    \item  \textbf{New applications and use cases:} Outside of traditional areas of healthcare and smart cities, the multimodal convergence of LLM and IoT should create new use cases in many domains. Imagine education spaces where IoT sensors recognize subtle student interest through nonverbal facial emotion and body signals and feed multimodal input to LLMs that dynamically change learning experiences in real time. In entertainment, consider immersive gaming, for example, where IoT sensors and wearables such as metaverse headset technologies combined with multimodal LLMs produce an adaptive story and a world that responds to player emotions and interactions \cite{kim2021multimodal}. 
        
    \item \textbf{Advancements in models:} The work in \cite{huang2024large} suggested that research should focus on minimizing the size of the model and the computational requirements. Novel architectures like Mamba offer optimistic solutions to the Transformer, where computational difficulties that are introduced by long sequences are being addressed. In addition, the paragraph recognizes the increasing power of in-context learning and prompt engineering as optimal ways to modify the model, potentially reducing the amount of computationally expensive fine-tuning needed \cite{lester2021power}. Finally, the focus is moving away from parameter size to emphasize data diversity and leverage larger, more heterogeneous datasets, foreshadowing the shift toward fewer large but more impactful models deeply rooted in rich and long pre-training, particularly in the domain of LLMs.

    \item \textbf{Quantum-enhanced models:} The authors in \cite{sarhaddi2025llms} highlighted the need to explore the integration of quantum computing to address the increasing complexity of LLMs in IoT environments driven by heterogeneous data. Although existing quantum-LLM research advances the enhancement of existing LLM architectures, a fundamental gap is the investigation of the explicit integration of quantum circuits into the IoT framework for LLM deployment \cite{liu2024towards}. Specifically, the research must look at how quantum computing can assist in deploying and integrating advanced LLMs in IoT setups, particularly for handling large datasets and complex models.
\end{enumerate}

\section{CONCLUSION}
\label{sec:conc}
In conclusion, this survey has opened up the vast potential of MLLMs being combined with the Internet of Things in 6G networks. This combination will be able to open new avenues in every industry, ranging from improving healthcare through improved remote care to revolutionizing agriculture through data-driven decision making and enabling smart cities in the truest sense. The key argument placed in this article is that by synergistic bundling of such technologies, we can go beyond the limits of current IoT deployments and leverage MLLMs' state-of-the-art multimodal perception and inference features. However, this vision can become a reality if serious consideration and future solutions are provided to a variety of problems. These include IoT hardware constraints of IoT devices, scalability and storage constraints of big multimodal data, and communication bottlenecks. Also, critical non-technical matters such as privacy of the data and security concerns, following in the footsteps of legal concerns, and user trust need to be treated properly so that deployment is not just appropriate, but ethical as well. Subsequent work should focus on developing new models and architectures that are computationally cheap and privacy-oriented, finding quantum-powered models, and developing new applications that most effectively utilize the multimodal synergy of 6G, IoT, and MLLMs to the maximum to provide the beginning to an entirely smart and networked world.

\section*{List of Acronyms}

\begin{table}[htbp]
\centering
\begin{tabular}{@{}l@{\hspace{0.5cm}}l@{}}
\toprule
\textbf{Acronym} & \textbf{Description} \\
\midrule
LLM & Large language model \\
IoT & Internet of Things \\
MLLM & Multimodal Language Models \\
LLM & Large language model \\
6G & 6th Generation Networks \\
AGI & Artificial General Intelligence \\
RNNs & Recurrent Neural Networks \\
CNNs & Convolutional Neural Networks \\
IMU & Inertial Measurement Unit \\
VLM & Vision Language Model \\
PALM-E & Pathways Language Model - Embodied \\
ADAS & Advanced Driver Assistance Systems \\
MEC & Mobile Edge Computing \\
QoS & Quality of Service \\
ISAC & Integrated Sensing and Communication \\
GPS & Global Positioning System \\
RGB-D & RGB and Depth \\
LiDAR & Light Detection and Ranging \\
PLS & Physical Layer Security \\
GNNs & Graph Neural Networks \\
BERT & Bidirectional Encoder Representations from Transformers \\
GPT-4 & Generative Pre-trained Transformer 4 \\
BART & Bidirectional and AutoRegressive Transformers \\
FL & Federated Learning \\
DP & Differential Privacy \\
eMBB & Enhanced Mobile Broadband \\
URLLC & Ultra-Reliable Low-Latency Communications \\
mMTC & massive Machine-Type Communications \\
AI & Artificial Intelligence \\
ML & Machine Learning \\
AR & Augmented Reality \\
XR & Extended Reality \\
VR & Virtual Reality \\
AIoT & AI and IoT \\
Tbps & Terabit per second \\
GHz & Gigahertz \\
ms & milliseconds \\
mURLLC & massive URLLC \\
mBRLLC & mobile Broadband Reliable Low-Latency Communication \\
IoMT & Internet of Medical Things \\
IIoT & Industrial Internet of Things \\
CPS-IoT & Cyber-Physical Systems IoT \\
\\
\bottomrule
\end{tabular}
\end{table}

\bibliographystyle{IEEEtran}
\bibliography{biblio}
\end{document}